\documentclass[superscriptaddress,runinaddress,showpacs,showkeys,aps,prd,floatfix,onecolumn]{revtex4}
\usepackage{amssymb,amsmath,epsfig}
\usepackage{amsmath,graphicx}
\usepackage{palatino}
\usepackage{changes}
\usepackage[colorlinks=true,
            linkcolor=blue,
            urlcolor=blue,
            citecolor=blue]{hyperref}
\begin{document}

\title{Weak gravitational lensing by Einstein-non-linear-Maxwell-Yukawa black hole}

\author{Wajiha Javed}
\email{wajiha.javed@ue.edu.pk; wajihajaved84@yahoo.com} 
\affiliation{Division of Science and Technology, University of Education, Township-Lahore, Pakistan}

\author{Muhammad Bilal Khadim}
\email{blaljutt723@gmail.com}
\affiliation{Division of Science and Technology, University of Education, Township-Lahore, Pakistan}

\author{Ali {\"O}vg{\"u}n}
\email{ ali.ovgun@emu.edu.tr}\homepage{https://aovgun.weebly.com} 

\affiliation{Physics Department, Eastern Mediterranean
University, 99628, Famagusta, North Cyprus via Mersin 10, Turkey.}

\begin{abstract}
    
In this article, we analyze the weak gravitational lensing in the
context of Einstein-non-linear Maxwell-Yukawa black hole.
To this desire, we derive the deflection angle of light by Einstein-non-linear Maxwell-Yukawa black hole
using the Gibbons and Werner method. For this purpose, we obtain the Gaussian curvature and apply the Gauss-Bonnet theorem to find
the deflection angle of Einstein-non-linear Maxwell-Yukawa black hole in weak field limits.
Moreover, we derive the deflection angle of light in the influence of plasma medium.
We also analyze the graphical behavior of deflection angle by Einstein-non-linear Maxwell-Yukawa black hole in
the presence of plasma as well as non-plasma medium.\end{abstract}


\date{\today}
 \maketitle
\section{Introduction}
In our universe the black holes (BHs) are essential components and
when stars are died, they become very dense objects which is the most important discoveries of astrophysics.
To test the fundamental laws of universe black holes provide a
golden opportunity. For example neutron star merges and the gravitational
waves from black holes has been discover recently. Therefore, the gravitational
lensing of black holes has attain incredible attention in the last few
decade, essentially because of the solid proof of supermassive
black holes at the focal point of galaxies \cite{1,2}. By utilizing the gravitational
lensing we can simplified the study of black holes, which is a general
investigative method for acquiring the time delays of the images, magnification
and positions. Darwin \cite{3} was the first who analyses the
Schwarzschild geometry. After that in $1985$ Herschel \cite{4} published
his similar article and after that many authors \cite{5,6},
extended this geometry to general spherically symmetric black holes and Reissner-Nordstrom
geometry. Kerr black holes \cite{7}-\cite{10} was also discussed for acquiring
the time delays of the images , magnification and positions by using gravitational
lensing. Many modification has been done through deflection of light
and modification in the reference of non-linear electrodynamics (NLE) \cite{11} has been
studied through the various alternative gravity theories \cite{12}. Classically,
a black hole contains singularity and also horizon because in general theory
of relativity the specetime singularities create a dozen of problems which are physical
as well as mathematical. Therefore many people use various methods to remove these
singularities from the black holes like modified gravity, effect of quantum
gravity and NLE. Bozza \cite{13} discussed these topics
in his recent article in details which also include observational prospectus and additional references \cite{14}-\cite{22}.

The main focus of this article is to calculate the deflection angle of photon by using GBT.
The aspect of deflection of light has been widely studied in different astrophysical
system in the influence of strong as well as weak gravitational lensing \cite{20}-\cite{23}. Gibbons
and Werner \cite{24} made a very important role who contended about the significance of
topological effect on the deflection of angle by utilizing the GBT and optical geometry.
Additionally, they have calculated deflection angle for the schwarzschild black holes which
is different from the other method by supposing light ray by taking domain outside, 
where the mass is closed in the given area on space
is strongly related to the lensing effect. Recently, Gibbons and Werner (GW) \cite{24}
computed the deflection angle by applying GBT as follows
 \begin{equation}
\sigma=-\int\int_{D\infty}\mathcal{K}dS.\nonumber\\
\end{equation}
Here, $\mathcal{K}$ is Gaussian curvature and $dS$ represents the surface element.
Afterwards, Werner \cite{25} extend this method and obtained the deflection of light by Kerr-Newman
BHs by applying the Nazims's method for Rander-Finsler metric. Recently,
deflection angle for a asymptotically flat and spherically symmetric spacetime
by using the finite distance from an observer to a light source has been calculated by Ishihara et al.
\cite{26}-\cite{28}. Moreover, Asada et al. \cite{29} in stationary
axisymmetric spacetime have calculated the weak gravitational lensing. In all of these
methods by using GBT they have calculated the weak gravitational lensing which shows the
global properties. The study of gravitational lensing in the presence of plasma
medium discussed in number of cases. Initially, Bisnovatyi-kogan \cite{30}-\cite{31}  show
that the gravitational deflection in plasma is different from vacuum deflection
angle due to presive properties of plasma. Recently, Gallo and Crisnejo \cite{32} discussed
the deflection angle of photon in a plasma medium.

In Einstein-Maxwell
theory (EMT) for hairy BH in the context of Weak gravitational lensing with a non-minimally coupled dilaton has been examined by Javed et al. \cite{33}. After that,
there is a lot of literature \cite{34}-\cite{Javed:2020fli} related to the investigation of weak gravitational lensing through the method of
GBT on various black holes, cosmic strings and wormholes. In this paper,
we study the weak gravitational lensing by Einstein-non-linear Maxwell-Yukawa black holes.

This paper is composed as; In Section 2, we concisely review about Einstein-non-linear Maxwell-Yukawa black hole.
In Section 3, by using the Gauss-Bonnet theorem we compute the deflection angle of Einstein-non-linear Maxwell-Yukawa black hole.
In Section 4, we work to investigate the deflection angle in the influence of plasma medium. We also demonstrate the graphical effect of deflection angle in the context of Einstein-non-linear Maxwell-Yukawa black
hole for plasma and also for non-plasma medium.
In Section 5, we present our result.
\section{Computation of weak lensing by Einstein-non-linear Maxwell-Yukawa black hole and Gauss-Bonnet theorem}
The Einstein-non-linear Maxwell-Yukawa BH in a static and spherically symmetric form is given as \cite{47}
\begin{equation}
ds^{2}=-f(r)dt^{2}+\frac{dr^{2}}{f(r)}+{r}^{2}(d\theta^{2}+\sin^{2}\theta d\phi^{2}), \label{space}
\end{equation}
The metric function $f(r)$ yields
\begin{equation}
f(r)\simeq 1+\frac{2M}{r}-\frac{4qC_{0}}{r^2}+\frac{4qC_{0}\alpha}{3r}-qC_{0}\alpha^{2}
+\mathcal{O}(\alpha^3). \label{f}\\
\end{equation}
Here, $M$ rendered as mass of BH, $C_{0}$
is an integration constant \cite{48}, $q$ represents charge of BH that is located
at the origin and $\alpha$ is a positive constant and it can be chosen as $\alpha=1$.\\

The optical spacetime simply written as
\begin{equation}
dt^{2}=\frac{dr^2}{ f(r)^{2}}+\frac{{r}^{2}d\phi^2}{ f(r)}
\end{equation}
From Eq. \ref{space} we obtain non-zero Christopher symbols given below
\begin{equation}
\Gamma^{0}_{00}=\frac{-f'({r})}{f({r})} ~,~ \Gamma^{0}_{11}=
\frac{2f({r})-{r}f'({r})}{2{r}f({r})}
~,~ \Gamma^{1}_{01}=\frac{{r}^2f'({r})-f({r})2{r}}{2}.\\
\end{equation}
Now, we calculate the Ricci Scalar corresponds to the optical metric by using
the non-zero Christopher symbols that are stated as:
\begin{equation}
\mathcal{R}=\frac{-f'({r})2}{2}+{f''({r})f({r})}.
\end{equation}
The Gaussian curvature that is computed as follows:
\begin{equation}
\mathcal{K}=\frac{RicciScalar}{2}\\
\end{equation}
After simplifying, Gaussian optical curvature for Einstein-non-linear Maxwell-Yukawa black hole is stated as:
\begin{equation}
\mathcal{K}=\frac{-f'({r})^2}{4}+\frac{f({r})f''(r)}{2},\label{AH6}
\end{equation}
where $f(r)$ is given in Eq. \ref{f}
so that Gaussian optical curvature for Einstein-non-linear Maxwell-Yukawa black hole in leading order term is obtained as;
\begin{eqnarray}
\mathcal{K}&\approx&-12\,{\frac {{ C_0}\,q}{{r}^{4}}}+4\,{\frac {{ C_0}\,q\alpha}{3{r
}^{3}}}+ \left( 2\,{r}^{-3}-24\,{\frac {{ C_0}\,q}{{r}^{5}}}+4\,{
\frac {{C_0}\,q\alpha}{{r}^{4}}} \right) M.\label{AH6}
\end{eqnarray}

\section{Deflection angle of Einstein-non-linear Maxwell-Yukawa black holes and Gauss-Bonnet theorem}
Now, by utilizing GBT, we will calculate the deflection angle of photon by Einstein-non-linear Maxwell-Yukawa BH.
By using GBT in the region $\mathcal{G}_{R}$, given as
\begin{equation}
\int_{\mathcal{G}_{R}}\mathcal{K}dS+\oint_{\partial\mathcal{G}_{R}}kdt
+\sum_{t}\hat{\alpha}=2\pi\mathcal{X}(\mathcal{G}_{R}),
\end{equation}
where, $k$ represent the geodesic curvature, $\mathcal{K}$ denotes
the Gaussian curvature and one can define $k$ as $k=\bar{g}(\nabla_{\hat{\alpha}}\hat{\alpha},\check{\alpha})$
in that way $\bar{g}(\hat{\alpha},\hat{\alpha})=1$, where $\hat{\alpha}$ represent
the unit acceleration vector and $\alpha_{t}$ denotes the exterior angle
at $t^{th}$ vertex respectively. As $R\rightarrow\infty$, we obtain the jump angles approximate to
$\pi/2$, Thus $\alpha_{O}+\alpha_{S}\rightarrow\pi$. Here, $\mathcal{X}(\mathcal{G}_{R})=1$ is a Euler
characteristic number and $\mathcal{G}_{R}$ denotes the non-singular domain. Therefore, we obtain
\begin{equation}
\int\int_{\mathcal{G}_{R}}\mathcal{K}dS+\oint_{\partial
\mathcal{G}_{R}}kdt+\hat{\alpha}=2\pi\mathcal{X}(\mathcal{G}_{R}).
\end{equation}
where, the total jump angle is $\hat{\alpha}=\pi$,
 When $R$ approching to infinity, then the remaining part is
$k(D_{R})=\mid\nabla_{\dot{D}_{R}}\dot{D}_{R}\mid$.
For geodesic curvature the radial component is described as:
\begin{equation}
(\nabla_{\dot{D}_{R}}\dot{D}_{R})^{r}=\dot{D}^{\phi}_{R}
\partial_{\phi}\dot{D}^{r}_{R}+\Gamma^{r}_{\phi\phi}(\dot{D}^{\phi}_{R})^{2}. \label{11}
\end{equation}
At $R$ very high , $D_{R}:=r(\phi)=R=const$. Thus, the first
term of Eq. \ref{11} vanishes and $(\dot{D}^{\phi}_{R})^{2}
=\frac{1}{f^2(r^\star)}$. Recalling $\Gamma^{r}_{\phi\phi}=
\frac{2f({r})-{r}f'({r})}{2{r}f({r})}$, we get
\begin{equation}
(\nabla_{\dot{D}^{r}_{R}}\dot{D}^{r}_{R})^{r}\rightarrow\frac{-1}{R},
\end{equation}
and  $k(D_{R})\rightarrow R^{-1}$, so we write $dt=Rd\phi$. Thus;
\begin{equation}
k(D_{R})dt=d\phi.
\end{equation}
From the pervious results, we get
\begin{equation}
\int\int_{\mathcal{G}_{R}}\mathcal{K}ds+\oint_{\partial\mathcal{G}_{R}}kdt
=^{R \rightarrow\infty }\int\int_{S_{\infty}}\mathcal{K}dS+\int^{\pi+\sigma}_{0}d\phi.\label{bilal2}
\end{equation}
At $0^{th}$ order weak field deflection limit of the light is defined as $r(t)=b/\sin\phi$.
Hence, the deflection angle given as:
\begin{equation}
\sigma=-\int^{\pi}_{0}\int^{\infty}_{b/\sin\phi}\mathcal{K}\sqrt{det\bar{g}}dud\phi,\label{AH7}
\end{equation}
here we put the first terms of Eq. \ref{AH6} into above Eq. \ref{AH7},
so we get the deflection angle upto leading order term is computed as:
\begin{eqnarray}
\sigma\simeq \frac{4M}{b}+{\frac {{ C_0}\,M\alpha\,q\pi}{{b}^{2}}}+8\,{\frac {{ C_0}\,q
\alpha}{3b}}-3\,{\frac {{ C_0}\,q\pi}{{b}^{2}}}-{\frac {32\,{ C_0}
\,Mq}{3\,{b}^{3}}}
. \label{alpha}
\end{eqnarray}

Note that the solution (\ref{alpha}) with $\alpha=q=0$ reduces to deflection angle of Schwarzcshild BH in the leading order terms. Moreover, it can be seen that the $\alpha$ parameter increases the deflection angle.

\section{Weak lensing by Einstein-non-linear Maxwell-Yukawa black hole in a plasma medium}

This section is based on the calculation of
weak gravitational lensing by Einstein-non-linear Maxwell-Yukawa black hole in plasma medium. For
Einstein non-linear Maxwell-Yukawa black hole the refractive index $n(r)$, is obtain as,
\begin{equation}
n(r)=\sqrt{1-\frac{\omega_{e}^{2}}{\omega_{\infty}^{2}}f(r)},
\end{equation}
where electron plasma
frequency is $\omega_{e}$ and photon frequency measured by an observer at
infinity is $\omega_{\infty}$ are  and  respectively, then the corresponding optical metric illustrated as
\begin{equation}
dt^{2}=g^{opt}_{ij}dx^{i}dx^{j}=n^{2}(r)\left[-f(r)dt^{2}+\frac{dr^{2}}{f(r)}+{r}^{2}(d\theta^{2})\right].
\end{equation}
The metric function $f(r)$ is defined in Eq. $2.4$ and
the non-zero christopher symbols corresponding to the optical metric are computed as:
\begin{eqnarray}
\Gamma^0_{00}&=&\left(1+\frac{\omega_e^2f({r})}{\omega_\infty^2}\right)
\left[\frac{-\omega_e^2f'({r})}{2\omega_\infty^2}\right]
-\frac{-f'({r})}{f({r})},\nonumber\\
\Gamma_{0 1}^{1}&=&\left(1+\frac{\omega_e^2f({r})}{\omega_\infty^2}\right)
\left(\frac{-\omega_e^2f'({r})}{2\omega_\infty^2}
\right)+\frac{2f({r})-{r}f'({r})}{2{r}f({r})},\nonumber\\
\Gamma^0_{11}&=&\left(1+\frac{\omega_e^2f({r})}{\omega_\infty^2}\right)
\left(\frac{\omega_e^2f({r})f'({r})r^2}{2\omega_\infty^2}\right)
+\frac{{r}^2f'({r})-f({r}){2r}}{2}.\nonumber\\
\end{eqnarray}
Now, by using the above non-zero christopher symbols the value of Gaussian optical curvature is found as:
\begin{eqnarray}
\mathcal{K}\approx 3\,{\frac {{\omega_e}^{2}M}{{r}^{3}{\omega_\infty}^{2}}}+2
\,{\frac {M}{{r}^{3}}}-104\,{\frac {{C_0}\,qM{\omega_e}^{2}
}{{\omega_\infty}^{2}{r}^{5}}}-24\,{\frac {{ C_0}\,qM}{{r}^{5}}}-
20\,{\frac {{C_0}\,q{\omega_e}^{2}}{{\omega_\infty}^{2}{
r}^{4}}}-12\,{\frac {{ C_0}\,q}{{r}^{4}}}+16\,{\frac {q\alpha\,{
C_0}\,{\omega_e}^{2}M}{{\omega_\infty}^{2}{r}^{4}}}+4\,{
\frac {q\alpha\,{ C_0}\,M}{{r}^{4}}}\\+2\,{\frac {q\alpha\,{\it C_0}\,{
\omega_e}^{2}}{{r}^{3}{\omega_\infty}^{2}}}+4/3\,{\frac {q
\alpha\,{ C_0}}{{r}^{3}}} \notag
.\label{AH61}
\end{eqnarray}
For this, we use GBT to compute the deflection
angle in order to relate it with non-plasma. As follows, for calculating
angle in the weak field limits, as the light beams become a straight line.
Therefore, the condition at zero order is $r=\frac{b}{sin\phi}$. Then the GBT reduces to this simple form for calculating the deflection angle $\sigma$:
\begin{equation}
\sigma=-\lim_{R\rightarrow 0}\int_{0} ^{\pi} \int_\frac{b}{\sin\phi} ^{R} \mathcal{K} dS
\end{equation}
So, the deflection angle of Einstein-non-linear  Maxwell-Yukawa BH in the presence of plasma medium is obtained as follows:
\begin{eqnarray} \label{AH677}
\sigma \simeq \frac{4M}{b}+4\,{\frac {q\alpha\,{ C_0}\,M{\omega_e}^{2}\pi}{{b}^{2}{
\omega_\infty}^{2}}}+{\frac {{ C_0}\,qM\alpha\,\pi}{{b}^{2}}}+4\,
{\frac {{ C_0}\,q\alpha\,{\omega_e}^{2}}{b{\omega_\infty}
^{2}}}+8/3\,{\frac {q{ C_0}\,\alpha}{b}}-5\,{\frac {{ C_0}\,q{
\omega_e}^{2}\pi}{{b}^{2}{\omega_\infty}^{2}}}-3\,{\frac {q
{C_0}\,\pi}{{b}^{2}}}-{\frac {416\,q{\it C_0}\,M{\omega_e}^{
2}}{9\,{b}^{3}{\omega_\infty}^{2}}}\\-{\frac {32\,{\it C_0}\,qM}{3\,{b
}^{3}}}+6\,{\frac {M{\omega_e}^{2}}{b{\omega_\infty}^{2}}} \notag.
\end{eqnarray}
The above result tells us that photon rays are moving into medium of
homogeneous plasma. One can see that the plasma effect can be removed
by neglecting $\frac{\omega_{e}}{\omega_{\infty}}$ term from Eq. \ref{AH677}
and it reduced into Eq. \ref{alpha}.

\section{Graphical influence of deflection angle on Einstein-non-linear Maxwell-Yukawa black hole}
This section of the paper comprises the graphical influence of deflection angle of Einstein-non-linear Maxwell-Yukawa BH.
We examine the impact of different parameters on deflection
angle. Here for simplicity we take ${\it C_0}=c$
\begin{center}
\epsfig{file=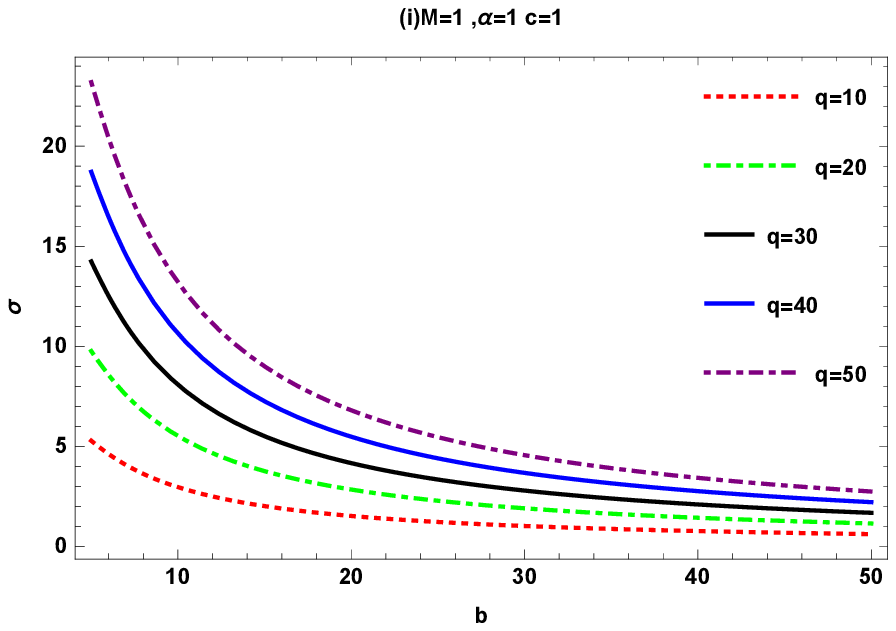,width=0.45\linewidth}\epsfig{file=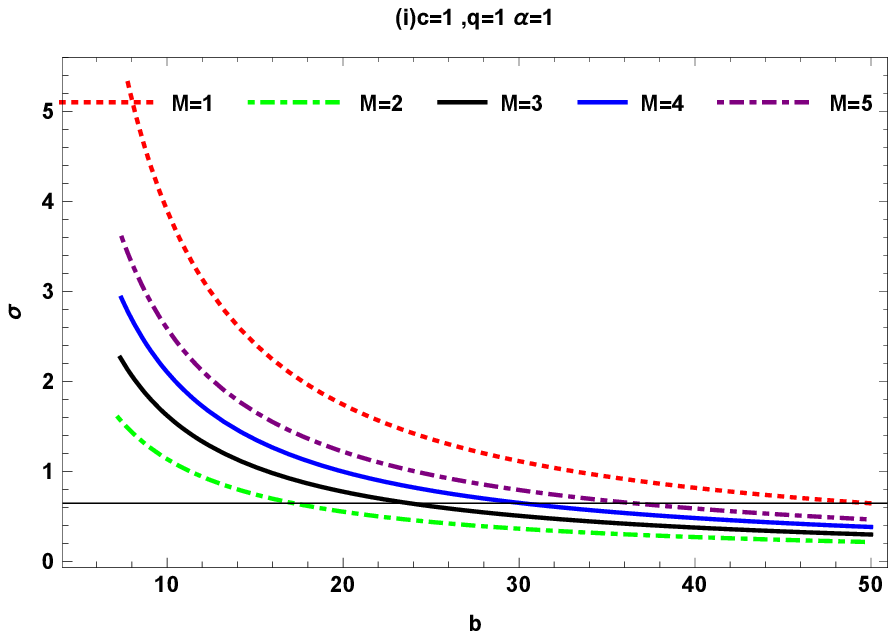,width=0.45\linewidth}\\
{Figure 1: $\sigma$ versus $b$}.\\ 
\end{center}
\begin{itemize}
\item  \textbf{Figure 1 Left Plot} indicates the behavior of deflection angle
w.r.t $b$ by fixing the value of $M$,$c$,$\alpha$ and varying $q$.
It is to be observed that for small constant value of $M$ and $q\geq0$ the behavior of deflection angle gradually decreasing with respect to impact parameter $b$. 
For increasing value
of $q$, deflection angle increases.
\end{itemize}

\begin{itemize}
\item  \textbf{Figure 1 Right Plot} represents the behavior of deflection angle w.r.t
$b$ by varying the mass $M$ and taking $q$,$c$,$\alpha$ fixed.
We noticed that for values of $M\geq0$ the behavior of deflection angle gradually
positively decreasing with respect to $b$. But for the increasing
values of $M$, the behavior of the deflection angle is increasing.
\end{itemize}
\begin{center}
\epsfig{file=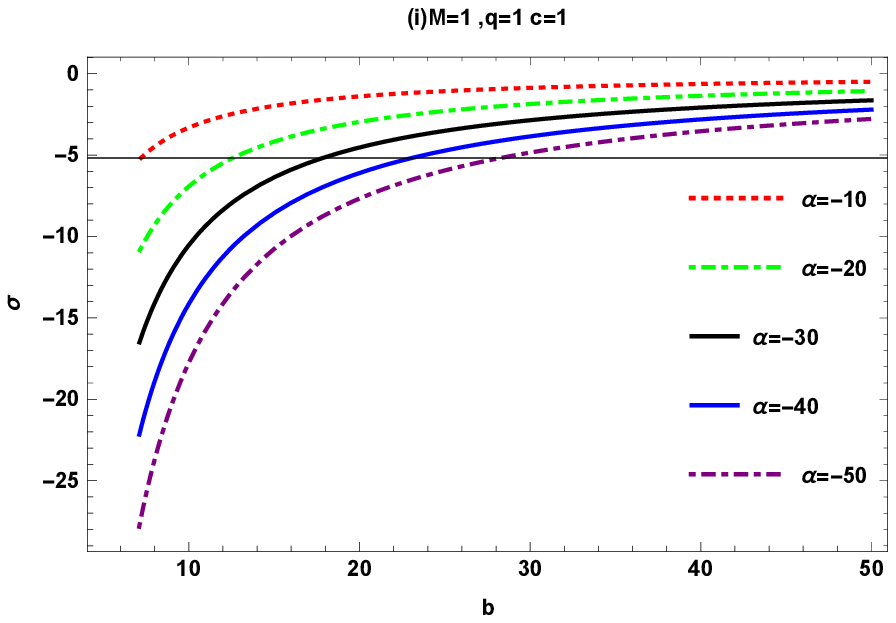,width=0.45\linewidth}\epsfig{file=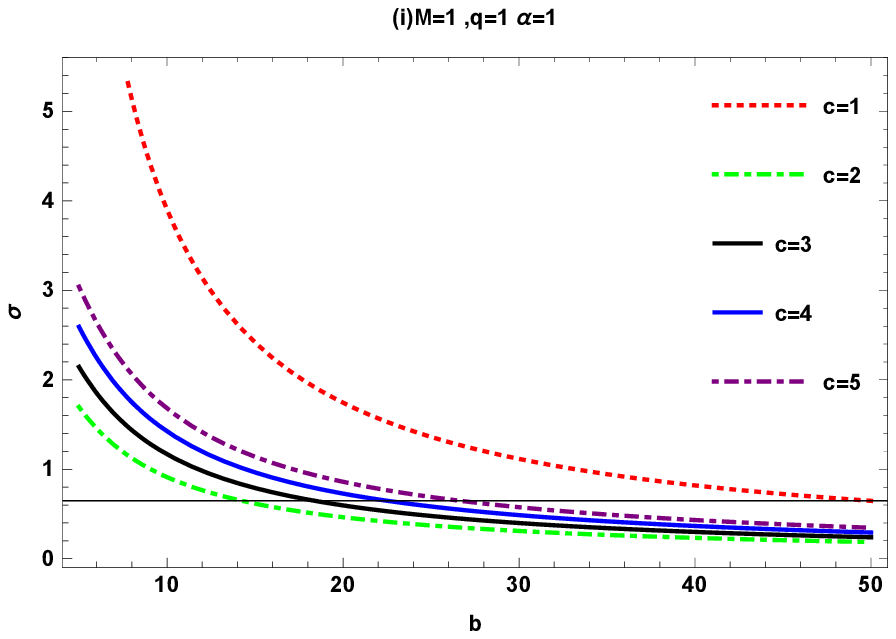,width=0.45\linewidth}\\
{Figure 1(a): $\sigma$ versus $b$}.\\
\end{center}
\begin{itemize}
\item  \textbf{Figure 1(a) Left Plot} indicates the behavior of deflection angle
w.r.t $b$ by fixing the value of $\alpha$ and varying $q$.$c$,$M$
It is to be observed that for decreasing the value of $\alpha<0$ the deflection angle gradually decreasing. On the other hand, the deflection angle is increasing gradually with respect to $b$ for fixed $\alpha$.
\end{itemize}
\begin{itemize}
\item  \textbf{Figure 1(a) Right Plot} represents the behavior of deflection angle w.r.t
$b$ by varying the $c$ and taking $q$,$\alpha$,$M$ fixed.
We noticed that for values of $b>5$ and $c\geq0$ the behavior of deflection angle gradually positively
decreasing. For the increasing the value of $c$,
the behavior of deflection angle is negatively decreasing.
\end{itemize}

\section{Graphical analysis for plasma medium}
This section gives us graphical analysis of deflection angle $\alpha$
in the presence of plasma medium. For simplicity we take ${\it C_0}=c$ in the Eq. \ref{AH677}.

\begin{center}
\epsfig{file=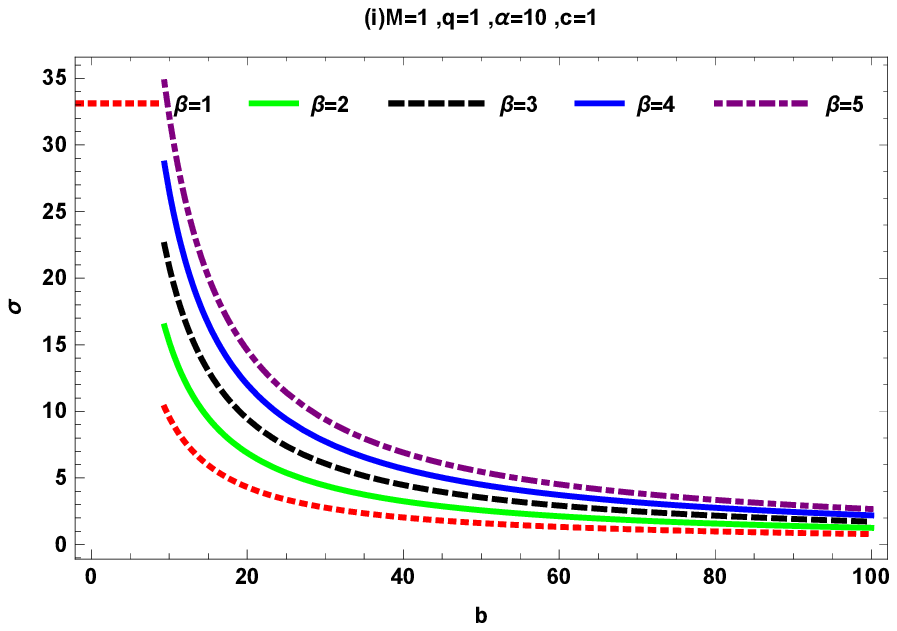,width=0.45\linewidth}\epsfig{file=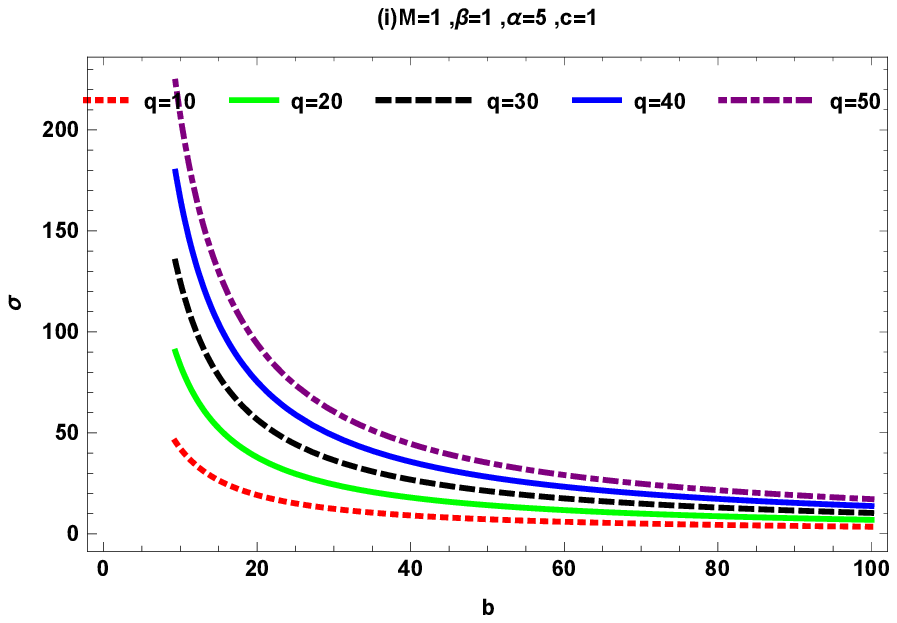,width=0.45\linewidth}\\
{Figure 2: $\sigma$ versus $b$}.\\
\end{center}
\begin{itemize}
\item  \textbf{Figure 2 Left Plot}, shows the behavior of deflection angle
w.r.t $b$ by taking the fixed values of $M$, $q$,$c$,$\alpha$ and by varying $\beta$.
It is to analyzed that initially the behavior of deflection angle is positively decreasing for $\beta\geq0$.
Furthermore, for the increasing value of $\beta$ the deflection angle positively increases for fixed $b$.
\end{itemize}
\begin{itemize}
\item  \textbf{Figure 2 Right Plot}, illustrate the behavior of deflection angle
w.r.t $b$ by varying BH charge $q$ and for fixed $M$,$\alpha$,$c$ and $\beta$.
We examined that for positive values of $q$ the deflection angle gradually positively decreasing. While, for
increasing values of $q$, the behavior of deflection angle is gradually
increasing for fixed $b$.
\end{itemize}
\begin{center}
\epsfig{file=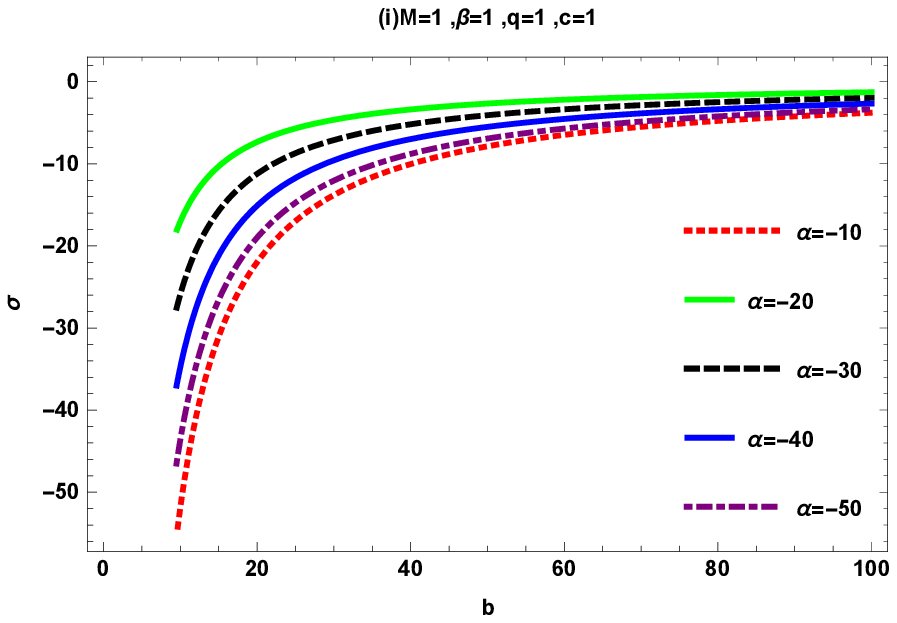,width=0.45\linewidth}\epsfig{file=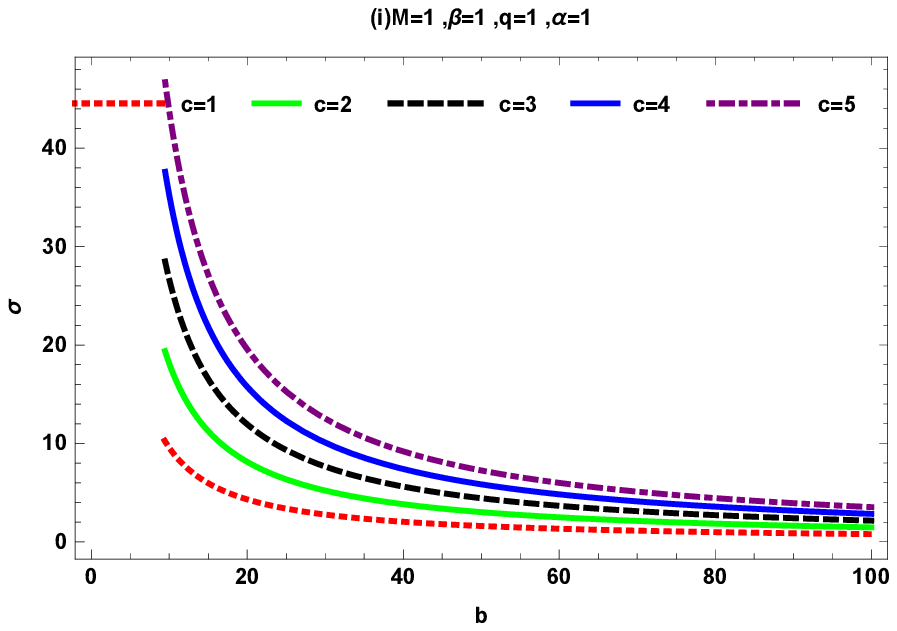,width=0.45\linewidth}\\
{Figure 2(a): $\tilde{\alpha}$ versus $b$}.\\
\end{center}
\begin{itemize}
\item  \textbf{Figure 2(a) Left Plot}, examines the behavior of deflection angle
w.r.t $b$ by taking the fixed values of $\beta$,$q$,$c$,$M$ and by varying $\alpha$.
We noticed that for small values of $\alpha$,and $\beta=1$, $q=1$
then the deflection angle cannot define the behavior. While, for
large values of $\alpha$ the behavior of deflection angle is positively decreasing.
For values of $\alpha<0$ the behavior of deflection angle gradually increasing.
\end{itemize}
\begin{itemize}
\item  \textbf{Figure 2(a) Right Plot}, we analyzed the behavior of deflection angle
w.r.t $b$ by varying $c$ and for fixed $M$,$\alpha$,$q$ and $\beta$. we see that for $c\geq0$ the deflection 
angle positively decreasing. For increasing values of $c$, deflection angle is increasing for fixed $b$.

\end{itemize}
\begin{center}
\epsfig{file=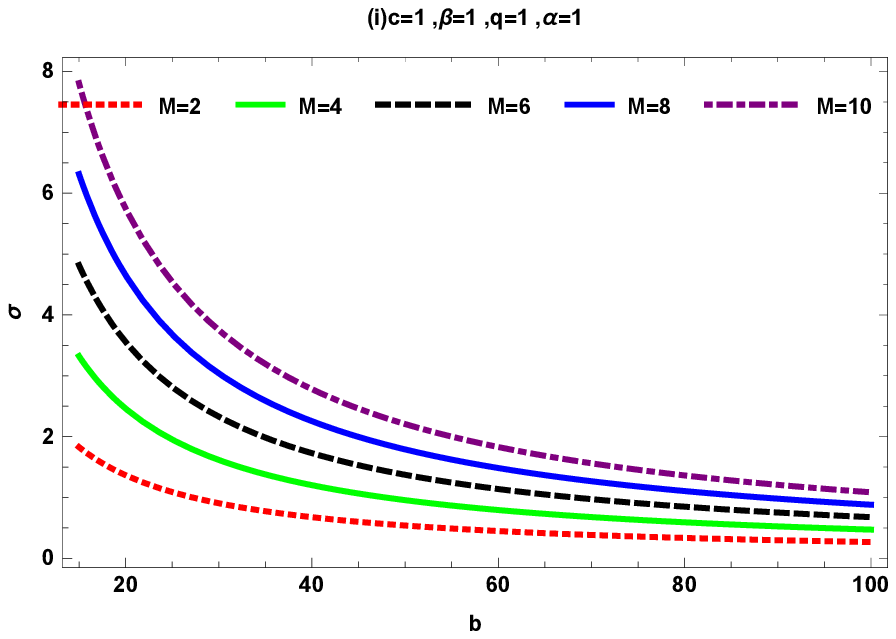,width=0.45\linewidth}\\
{Figure 3: $\sigma$ versus $b$}.\\
\end{center}
\begin{itemize}
\item  \textbf{Figure 3}, manifest the influence of deflection
angle w.r.t $b$ for constant values of $q$, $\beta$,$c$,$\alpha$ and varying $M$.
It has been noted that for small range of $b<15$ the behavior of deflection angle not defined. On the other hand for large $b\geq15$ and for $M\geq0$ the deflection angle positively decreasing
and the behavior of deflection angle is increasing for
increasing values of $M$ at fixed $b$.
\end{itemize}

\section{Conclusion}
The present paper is about the investigation of deflection
angle by Einstein-non-linear Maxwell-Yukawa BH in non-plasma as well as plasma medium.
In this regard, we study the weak gravitational lensing by using
GBT and obtain the deflection angle of photon for Einstein-non-linear Maxwell-Yukawa BH.
The obtained deflection angle is given in Eq. \ref{alpha} as follows:
\begin{eqnarray}
\sigma\simeq \frac{4M}{b}+{\frac {{ C_0}\,M\alpha\,q\pi}{{b}^{2}}}+8\,{\frac {{ C_0}\,q
\alpha}{3b}}-3\,{\frac {{ C_0}\,q\pi}{{b}^{2}}}-{\frac {32\,{ C_0}
\,Mq}{3\,{b}^{3}}}+\mathcal{O}(q^2,M^2,C_0^2).\nonumber
\end{eqnarray}
We examine that by the reduction of some parameters the obtained deflection angle converted
into the Schwarzschild deflection angle up to the first order terms. We also discuss the graphical effect
of different parameters on deflection angle by Einstein-non-linear Maxwell-Yukawa BH. We also observed that deflection angle in the presence of plasma medium given by Eq. \ref{AH677} which is;
\begin{eqnarray}
\sigma \simeq \frac{4M}{b}+4\,{\frac {q\alpha\,{ C_0}\,M{\omega_e}^{2}\pi}{{b}^{2}{
\omega_\infty}^{2}}}+{\frac {{ C_0}\,qM\alpha\,\pi}{{b}^{2}}}+4\,
{\frac {{ C_0}\,q\alpha\,{\omega_e}^{2}}{b{\omega_\infty}
^{2}}}+8/3\,{\frac {q{ C_0}\,\alpha}{b}}-5\,{\frac {{ C_0}\,q{
\omega_e}^{2}\pi}{{b}^{2}{\omega_\infty}^{2}}}-3\,{\frac {q
{C_0}\,\pi}{{b}^{2}}}-{\frac {416\,q{\it C_0}\,M{\omega_e}^{
2}}{9\,{b}^{3}{\omega_\infty}^{2}}}\\-{\frac {32\,{\it C_0}\,qM}{3\,{b
}^{3}}}+6\,{\frac {M{\omega_e}^{2}}{b{\omega_\infty}^{2}}}+\mathcal{O}(q^2,M^2,C_0^2,\omega_e^3) \notag
\end{eqnarray}
When the $\frac{\omega_{e}}{\omega_{\infty}}$ approaches to zero,the plasma effect are removed.
Furthermore, we scrutinized the graphical impact of deflection
angle on Einstein-non-linear Maxwell-Yukawa BH in plasma medium versus some parameters. One can conclude that the deflection of photon is found outside of the lensing area which points that the gravitational lensing effect is a global and even topological effect. Hence, it is our belief that studying on the Virbhadra-Ellis lens equation, can provide us to 
improve the computation of the image positions, Einstein ring radii, magnification factors and the magnification ratio. This is going to be our
next problem in the near future. Moreover, it will be interesting to investigate the deflection angle of black holes in MAXWELL– $f(R)$ gravity theories  and in fourth order gravity theories using the GBT in future \cite{Nashed:2019tuk,Horvath:2012kf,Capozziello:2006dp}.

\end{document}